\documentclass[10pt, a4paper]{article}
\usepackage[hmarginratio=2:3]{geometry}

\usepackage{amssymb}
\usepackage{amsthm}

\usepackage{amsmath, calc, graphicx}
\usepackage[authoryear]{natbib}

\usepackage{lineno}

\usepackage{subfigure}
\usepackage{url}%

\theoremstyle{plain}
\newtheorem{theorem}{Theorem}[section]
\newtheorem{corollary}[theorem]{Corollary}

\newtheorem{proposition}[theorem]{Proposition}

\theoremstyle{definition}

\theoremstyle{remark}
\newtheorem{remark}[theorem]{Remark}
\newtheorem{example}[theorem]{Example}

\newcommand{\Pit}[1]{\mathbb{P} \! \left[#1\right]}

\newcommand{\Eit}[1]{\mathbb{E} \! \left[#1\right]}

\newcommand{\Var}[1]{\mbox{\rm Var} \! \left[#1\right]}

\begin{document}



\title{A Sheet of Maple to Compute Second-Order Edgeworth Expansions and Related Quantities of any Function of the Mean of an iid Sample of an Absolutely Continuous Distribution}

\author{F. Bertrand$^{\rm a}$$^{\ast}$\thanks{$^\ast$Corresponding author. Email: Frederic.Bertrand@math.unistra.fr
\vspace{6pt}} and M. Maumy-Bertrand$^{\rm a}$\\\vspace{6pt}  $^{a}${\em Institut de Recherche Math\'ematique Avanc\'ee,}\\ {\em Universit\'e de Strasbourg,}\\ {\em 7, rue Ren\'e Descartes, 67084 Strasbourg Cedex, France}}

\maketitle

\begin{abstract}
We designed a completely automated {\tt Maple} ($\geqslant 15$) worksheet for deriving Edgeworth and Cornish-Fisher expansions as well as the acceleration constant of the bootstrap bias-corrected and accelerated technique. It is valid for non-parametric or parametric bootstrap, of any (studentized) statistics that is -a regular enough- function of the mean of an iid sample of an absolutely continuous distribution.

This worksheet allowed us to point out one error in the second-order Cornish-Fisher expansion of the studentized mean stated in Theorem 13.5 by Das Gupta in \cite[p.~194]{dasg} as well as lay the stress on the influence of the slight change of the normalizing constant when computing the second-order Edgeworth and Cornish-Fisher expansions of the $t$-distribution as stated in Theorem 11.4.2 by   Lehman and Romano in \cite[p.~460]{lero}.

In addition, we successfully applied the worksheet to a complex maximum likelihood estimator as a first step to derive more accurate confidence intervals in order to enhance quality controls.

The worksheet also features export of {\tt Maple} results into {\tt R} code. In addition, we provide {\tt R} code to plot these expansions as well as their increasing rearrangements. All these supplemental materials are available upon request.



\end{abstract}

\section{Introduction}

Even though the Edgeworth expansion is often, at the first glance, mostly viewed as a theoretical tool, it has steadily gained interest from the practitioner point of view for several decades. At the time present, there are too many examples of smart applications of these expansions to list them all. For instance, it has successfully been used to improve confidence intervals, either asymptotic or bootstrap ones, giving sharp insights on, among other properties, their coverage or expected length. Hall \cite{hall92} and Shao and Tu \cite{jd96} provide a detailed account on this topic. Moreover, the Edgeworth expansion is deeply rooted in one of the more recent outstanding study dealing with confidence interval for a binomial proportion (Brown, Cai, and Das Gupta, \cite{bcd01,bcd02}), in an exponential family (Brown, Cai, and Das Gupta, \cite{bcd03}) and in discrete distributions (Cai, \cite{cai05}).

The major drawback of these expansions are their long and tedious computations. Even when dealing with the most simple statistic -the studentized mean-, a famous textbook can provide an incorrect formula as the second-order Cornish-Fisher expansion stated in Theorem 13.5 by Das Gupta in \cite[p.~194]{dasg}. Make a slight change in the denominator of the usual raw variance estimator to use instead its unbiased counterpart -by simply replacing $1/\sqrt{n}$ by $1/\sqrt{n-1}$ to get an exactly $t$-distributed statistic in the Gaussian framework- and it leads to different second-order formulas, which accounts for the discrepancies of the Edgeworth expansions for the studentized mean when compared to the Edgeworth expansion of the $t$-distribution, for instance as stated in Theorem 11.4.2 by Lehman and Romano in \cite[p.~460]{lero}.

 For a smooth -or regular enough- model $A(\overline{X})$ of the mean $\overline{X}$ of an iid sample $(X_1,\ldots,X_n)$ drawn from an absolutely continuous distribution $\mathcal{L}_{\theta}$, first and second-order Edgeworth expansions are valid. Closed, yet complex, formulas were derived by Withers \cite{with83} and Hall \cite{hall92} to compute them from the partial derivatives of $A$ and the central moments of the underlying distribution $\mathcal{L}_{\theta}$. In addition, in such a model that exhibits a valid second-order Edgeworth expansion, there exists a closed formula for the acceleration constant $a$ used in the bootstrap bias-corrected and accelerated technique (BCA). Were we able to evaluate this formula, this would help to remedy to ``A disadvantage of the bootstrap BCA'', as stated Shao and Tu \cite[p.~140]{jd96},  which is ``that the determination of $a$ is not easy, especially when the problem under consideration is complex''. Moreover, in some statistical models, as put forward in two cases in Sections~\ref{expvarian} and~\ref{morecomp}, the acceleration constant $a$ is free of the unknown parameters of the underlying distribution and thus can be derived exactly.

All these reasons account for our use of the {\tt Maple} computation engine \cite{Maplesoft}, that combines numeric computations with symbolic capabilities, in order to design a tool able to automatically and trustworthy compute Edgeworth expansions as well as Cornish-Fisher ones or the acceleration constant $a$ used in BCA bootstrap technique. This is this tool, that we provide as a {\tt Maple} worksheet.

{\tt Maple} functions or expressions code can be translated into {\tt MATLAB} code, which in our setting, is virtually {\tt R} \cite{Rsoft} code. The {\tt R} language can then be used for display purposes, as we did with Figures~\ref{1} and~\ref{2}, to apply bootstrap techniques of the {\tt boot} package \cite{Pkgboot,dahi97} or for any further statistical analysis, for instance to deal with the non-increasing cumulative distribution functions that may result from the above Edgeworth expansions using the rearrangement operator \cite{hlp52} implemented in the {\tt Rearrangement} package \cite{cfg10,rearrR}. Using \cite{with83}, it is easy to derive Cornish-Fisher expansions of any higher order and upgrade our worksheet to have {\tt Maple} compute them.

Previous work by Finner and Dickhaus \cite{fd10} only addresses the part on the computation of Edgeworth expansions up to any order. From a (applied) statistical point of view, it is not very useful. On the contrary, we not only carry out the Edgeworth expansions but also the Cornish-Fisher ones, BCA acceleration constant derivation, export to and import in {\tt R} of the results, then use of the {\tt R} language to derive increasing rearrangements of the computed expansions, which sums up to a significant additional value of this article for statisticians and practitioners. 

\section{Edgeworth-Cornish-Fisher expansions and BCA bootstrap}\label{sect1}
\subsection{Introducing the Edgeworth expansion}
Let $\{\boldsymbol{Z_n}: n\geqslant 1\}$ be a sequence of random variables with 0 mean and unit variance. The cumulant generating function $\Psi_n$ of $\boldsymbol{Z_n}$ is defined by: $\Psi_n(t)=\log \Eit{it\boldsymbol{Z_n}}$, $t \in \mathbb{R}$. The $\nu$\textsuperscript{th} cumulant of $\boldsymbol{\boldsymbol{Z_n}}$ is given by: $\kappa_{\nu,n}=\frac{1}{i^{\nu}}\frac{d^{\nu}}{dt^{\nu}}\Psi(0)$, $\nu\in \mathbb{N}$.
Consequently, we show that:
\begin{eqnarray}
\label{E:3.1}
\Eit{\exp(it\boldsymbol{Z_n})}&=&\exp\left(\sum_{j=1}^{+\infty}\kappa_{j,n}\frac{1}{j!}(it)^{j}\right), \quad t\in\mathbb{R}.
\end{eqnarray}
The $\nu$\textsuperscript{th} derivative of the characteristic function of $\boldsymbol{Z_n}$ exists at the point $t=0$ if and only if $\Eit{|\boldsymbol{Z_n}|^{\nu}}<+\infty$. 
Then the $\nu$\textsuperscript{th} cumulant of $\boldsymbol{Z_n}$ exists if and only if $\Eit{|\boldsymbol{Z_n}|^{\nu}}<+\infty$.
The entire series expansion and the identification of the coefficients lead to the expression of the first four cumulants, see \cite{hall92} for more details:
\begin{eqnarray*}
\kappa_{1,n}=\Eit{\boldsymbol{Z_n}},&&
\kappa_{2,n}=\Var{\boldsymbol{Z_n}},\nonumber\\
\kappa_{3,n}=\Eit{\boldsymbol{Z_n}-\Eit{\boldsymbol{Z_n}}}^3,&&
\kappa_{4,n}=\Eit{\boldsymbol{Z_n}-\Eit{\boldsymbol{Z_n}}}^4-3\left(\Var{\boldsymbol{Z_n}}\right)^2.\nonumber
\end{eqnarray*}
Assume that there exist four real numbers $k_{1,2}$, $k_{2,2}$, $k_{3,1}$, $k_{4,1}$ so that the first four cumulants satisfy the following asymptotic expansions:
\begin{eqnarray*}
\kappa_{1,n}=n^{-1/2} k_{1,2}+O(n^{-3/2}),&&
\kappa_{2,n}=1+n^{-1}k_{2,2}+O(n^{-3/2}),\\
\kappa_{3,n}=n^{-1/2}k_{3,1}+O(n^{-3/2}),&&
\kappa_{4,n}=n^{-1}k_{4,1}+O(n^{-3/2}).
\end{eqnarray*}
Then Equation~\ref{E:3.1} becomes:
\begin{eqnarray*}
\Eit{\exp{(it\boldsymbol{Z_n})}}
&=&\exp{\left(-\frac{1}{2}t^2\right)}\Bigg\{1+ n^{-1/2}\left(k_{1,2}it+\frac{1}{6}k_{3,1}(it)^3\right)\\
&&
+n^{-1}\left(\frac{1}{2}k_{2,2}(it)^2+\frac{1}{24}k_{4,1}(it)^4
+\frac{1}{2}\left(k_{1,2}it+\frac{1}{6}k_{3,1}(it)^3\right)^2
\right)+\cdots\Bigg\}.\nonumber
\end{eqnarray*}
A Fourier inversion of this expansion leads to: 
\[\label{E:4.4}
\Pit{\boldsymbol{\boldsymbol{Z_n}}\leqslant x}=\Phi(x)+\left(n^{-1/2}p_1(x)+n^{-1}p_2(x)+\cdots\right)\phi(x),\nonumber
\]
uniformly in $x\in\mathbb{R}$ where $\phi(x)=\Phi'(x)$ is the probability density function of a standard Gaussian distribution and
\begin{eqnarray}
p_1(x)&=&-\left(k_{1,2}+\frac{k_{3,1}}{6}(x^2-1)\right),\label{eqp_{1}}\\
p_2(x)&=&-x\left(\frac{\left(k_{2,2}+k^2_{1,2}\right)}{2}
+\frac{\left(k_{4,1}+4k_{1,2}k_{3,1}\right)}{24}(x^2-3)+\frac{k^2_{3,1}}{72}(x^4-10x^2+15)\right).\label{eqp_{2}}
\end{eqnarray} 

\subsection{A model valid for Edgeworth expansions}\label{subsec12}
Let $\boldsymbol{X}$, $\boldsymbol{X_1}$, $\boldsymbol{X_2}$, \dots be independent and identically distributed random column $d$-vectors with mean $\mathbb{E}\left[\bf{X}\right]$ denoted by $\boldsymbol{\mu}$, and $\boldsymbol{\overline{X}}=n^{-1}\sum_{i=1}^n{\boldsymbol{X_i}}$. Let $A \ : \ \mathbb{R}^d \rightarrow \mathbb{R}$ be a smooth function satisfying $A(\boldsymbol{\mu})=0$. Denote the $i$th element of a $d$-vector $\boldsymbol{v}$ by: $v^{(i)}$ or $(\boldsymbol{v})^{(i)}$, 
\begin{eqnarray}
&a_{i_1\ldots i_j}=\left(\partial^j/\partial x^{(i_1)}\ldots\partial x^{(i_j)}\right)A(\mathbf{X})|_{\bf{X}=\boldsymbol{\mu}}&\\&\quad \mbox{and}\quad\nonumber&\\ 
&\mu_{i_1\ldots i_j}=\mathbb{E}\left[\left(\mathbf{X}-\boldsymbol{\mu}\right)^{(i_1)}\ldots\left(\mathbf{X}-\boldsymbol{\mu}\right)^{(i_j)}\right].&\label{defamu}
\end{eqnarray}
We will focus on functions such as:
\begin{enumerate}
	\item $A(\boldsymbol{x})=(g(\boldsymbol{x})-g(\boldsymbol{\mu}))/h(\boldsymbol{\mu})$, where $\theta_0=g(\boldsymbol{\mu})$ is the (scalar) parameter estimated by $\hat\theta=g(\boldsymbol{\overline{X}})$ and $h(\boldsymbol{\mu})^2$ is the asymptotic variance $\sigma_{\hat\theta}^2$ of $n^{1/2}\hat\theta$;
	\item $A(\boldsymbol{x})=(g(\boldsymbol{x})-g(\boldsymbol{\mu}))/h(\boldsymbol{x})$, where $h(\boldsymbol{\overline{X}})$ is an estimate of $h(\boldsymbol{\mu})=\sigma_{\hat\theta}$.
\end{enumerate}
We can assume that $h$ is a known function; as stated in \citep{hall92}, the following $h$ function fulfills the previous requirements: 
$h^2(\boldsymbol{x})=\sum_{i=1}^d\sum_{j=1}^d \left(\partial g/\partial x^{(i)}\partial g/\partial x^{(j)}\right)(\boldsymbol{x})\mu_{i,j}$. If used, $A_0$ stands for the non-studentized case and $A_s$ the studentized one. This ``smooth function model'' \citep{hall92} allows to study problems where $\theta_0$ is a mean, or a variance, or a ratio of means or variances, or a difference of means or variances, or a correlation coefficient, etc. 

\begin{example}\label{exempl1}
Let $\{W_1, \dots, W_n\}$ be a random sample from an univariate population with mean $m$ and variance $\beta^2$, and if we wished to estimate $\theta_0=m$, then we would take $d=2$, $\boldsymbol{X}=(X^{(1)},X^{(2)})^{\top}=(W,W^2)^{\top}$, $\boldsymbol{\mu}=\Eit{\boldsymbol{X}}$, $g(x^{(1)},x^{(2)})=x^{(1)}, \quad h^2(x^{(1)},x^{(2)})=x^{(2)}-(x^{(1)})^2$. This would ensure that $g(\boldsymbol{\mu})=m$, $g(\boldsymbol{\overline{X}})=\overline{W}$ (the sample mean), $h(\boldsymbol{\mu})=\beta^2$, and
\[
h^2(\boldsymbol{\overline{X}})=n^{-1}\sum_{i=1}^nX_i^{(2)}-\left(n^{-1}\sum_{i=1}^nX_i^{(1)}\right)^2=n^{-1}\sum_{i=1}^n(W_i-\overline{W})^2=\hat\beta^2 \ \textup{(the sample variance).}
\]
\end{example}
\begin{example}\label{exempl2}
Let $\{W_1, \dots, W_n\}$ be a random sample from an univariate population with mean $m$ and variance $\beta^2$, and if we wished to estimate $\theta_0=\beta^2$, then we would take $d=4$, $\boldsymbol{X}=(W,W^2,W^3,W^4)^{\top}$, $\boldsymbol{\mu}=\Eit{\boldsymbol{X}}$, $g(x^{(1)},x^{(2)},x^{(3)},x^{(4)})=x^{(2)}-(x^{(1)})^2$, $h(x^{(1)},x^{(2)},x^{(3)},x^{(4)})=x^{(4)}-4x^{(1)}x^{(3)}+6(x^{(1)})^2x^{(2)}-3(x^{(1)})^4-(x^{(2)}-(x^{(1)})^2)^2$.
In this case, we have:
\[
g(\boldsymbol{\mu})=\beta^2, \quad g(\boldsymbol{\overline{X}})=\hat\beta^2, \quad h(\boldsymbol{\mu})=\Eit{W-m}^4-\beta^4, \quad h(\boldsymbol{\overline{X}})=n^{-1}\sum_{i=1}^n(W_i-\overline{W})^4-\hat\beta^4.
\]
\end{example}
The cases where $\theta_0$ is a correlation coefficient (a function of five means), or a variance ratio (a function of four means), among others, may be treated similarly.

Since $A(\boldsymbol{\mu})=0$ and $n^{1/2}(\boldsymbol{\overline{X}}-\boldsymbol{\mu})=O_\mathbb{P}(1)$, by a Taylor expansion: $S_n=n^{1/2}A(\boldsymbol{\overline{X}})=S_{nr}+O_\mathbb{P}(n^{-r/2})$, $r\geqslant 1$, where $S_n-S_{nr}=O_\mathbb{P}(n^{-r/2})$ means that $\lim_{\lambda\to+\infty}\limsup_{n\to+\infty}\mathbb{P}({\left|S_n-S_{nr}\right|}$ ${/n^{-r/2}}>\lambda)=0$ and 
\[
S_{nr}=\sum_{i=1}^da_iZ^{(i)}+\frac{n^{-1/2}}{2}\sum_{i_1=1}^d\sum_{i_2=1}^da_{i_1i_2}Z^{(i_1)}Z^{(i_2)}+\cdots+\frac{n^{-(r-1)/2}}{r!}\sum_{i_1=1}^d\cdots\sum_{i_r=1}^da_{i_1\ldots i_r}Z^{(i_1)}\cdots Z^{(i_r)}.\nonumber
\]

\begin{theorem}[Hall \cite{hall92}, Bertrand \cite{Bertrand}]\label{T:5.1}
Assume that the function $A$ has $4$ continuous derivatives in a neighborhood of $\boldsymbol{\mu}=\Eit{\boldsymbol{X}}$, that $A(\boldsymbol{\mu})=0$, that $\Eit{\left\|\boldsymbol{X}\right\|^{4}}<+\infty$, and that the characteristic function $\chi$ of $\boldsymbol{X}$ satisfies
$\limsup_{\left\|\boldsymbol{t}\right\|\to+\infty}{\left|\chi(\boldsymbol{t})\right|}<1$. We have:
\begin{eqnarray}
\label{E:4.7}\Pit{\sqrt n A(\boldsymbol{\overline{X}})\leqslant x}&=&\Phi(x)+n^{-1/2}p_{1}(x)\phi(x)+n^{-1}p_{2}(x)\phi(x)+  \,O\left(n^{-3/2}\right),
\end{eqnarray}
uniformly in $x\in\mathbb{R}$ and where $p_{1}$, $p_{2}$ are defined by Equations~\ref{eqp_{1}} and~\ref{eqp_{2}}, the values of the coefficients $k_{1,2}$, $k_{2,2}$, $k_{3,1}$ et $k_{4,1}$ been given by:
\allowdisplaybreaks
\begin{eqnarray*}
k_{1,2}&=&\frac{1}{2}\sum_{i=1}^d\sum_{j=1}^da_{ij}\mu_{ij},\\
k_{2,2}&=&\sum_{i=1}^d\sum_{j=1}^d\sum_{k=1}^da_ia_{jk}\mu_{ijk}+\frac{1}{2}\sum_{i=1}^d\sum_{j=1}^d\sum_{k=1}^d\sum_{l=1}^d a_{ij}a_{kl}\mu_{ik}\mu_{jl}\\
&&+\sum_{i=1}^d\sum_{j=1}^d\sum_{k=1}^d\sum_{l=1}^d a_i a_{jkl}\mu_{ij}\mu_{kl},\\
k_{3,1}&=&\sum_{i=1}^d\sum_{j=1}^d\sum_{k=1}^d a_ia_ja_k \mu_{ijk}+3\sum_{i=1}^d\sum_{j=1}^d\sum_{k=1}^d\sum_{l=1}^da_ia_ja_{kl}\mu_{ik}\mu_{jl},\\
k_{4,1}&=&\sum_{i=1}^d\sum_{j=1}^d\sum_{k=1}^d\sum_{l=1}^da_ia_ja_ka_l\left(\mu_{ijkl}-3\mu_{ij}\mu_{kl}\right)\\
&&+12\sum_{i=1}^d\sum_{j=1}^d\sum_{k=1}^d\sum_{l=1}^d\sum_{m=1}^d a_ia_ja_ka_{lm}\mu_{il}\mu_{jkm}\\
&&+12\sum_{i=1}^d\sum_{j=1}^d\sum_{k=1}^d\sum_{l=1}^d\sum_{m=1}^d\sum_{o=1}^d a_ia_ja_{kl} a_{mo}\mu_{ik}\mu_{jm}\mu_{lo}\\
&&+4\sum_{i=1}^d\sum_{j=1}^d\sum_{k=1}^d\sum_{l=1}^d\sum_{m=1}^d\sum_{o=1}^d a_ia_ja_k a_{lmo}\mu_{il}\mu_{jm}\mu_{ko}.\label{eqkis}
\end{eqnarray*}
\end{theorem}
Using \cite{hall92} and \cite{with83}, it is easy to derive Edgeworth expansions of any higher order and complete our worksheet to have {\tt Maple} compute them. One can also try to use Finner and Dickhaus work \cite{fd10}.

\begin{remark}\label{rematdist}
Edgeworth expansions can also be used to derive asymptotic expansions for cumulative density functions of distribution. A case in point is the $t$-distribution, see Lehman and Romano in Formulas 11.73 and 11.74 of Theorem 11.4.2, \cite[p.~460]{lero} for the second-order Edgeworth expansion of the $t$-distribution. An easy way to retrieve an exactly $t$ distributed statistic in a iid $(X_1,\ldots,X_n)$ Gaussian $\mathcal{N}(\mu,\sigma)$ framework, is to consider $\sqrt{n} (\overline{X}-\mu) / \hat\sigma_c$ with $\hat\sigma_c$ the square root of the unbiased variance estimator ($1/(n-1)\sum_{i=1}^n(X_i-\overline{X})^2$). Unfortunately, when applying Theorem \ref{T:5.1} to the sample mean, we get the expansion of $\sqrt{n}( \overline{X}-\mu) / \hat\sigma$ with $\hat\sigma$ the square root of the raw variance estimator ($1/n\sum_{i=1}^n(X_i-\overline{X})^2$). This subtle change in the normalizing constant leads to discrepancies between the two second-order terms of the Edgeworth expansion of the $t$-distribution and of the studentized mean. Using asymptotic expansions, we show in the provided Maple worksheet, and report in Section~\ref{stumeansubsec}, how to switch from the expansion of the studentized mean to the one of the $t$-distribution.
\end{remark}

\subsection{Cornish-Fisher expansion}
The Cornish-Fisher expansion is the expansion of the $\alpha$-level quantile of an estimator from its Edgeworth expansion:
\[
w_{\alpha}=\inf\left\{x:\mathbb{P}\left[\sqrt nA(\boldsymbol{\overline{X}})\leqslant x\right]\geqslant\alpha\right\}.
\] 
Denote by $z_{\alpha}$ the $\alpha$ level quantile of a standard Gaussian distribution. Equation~\ref{E:4.7} leads to an expansion of $w_{\alpha}$ as a polynomial in $z_{\alpha}$:
\[
w_{\alpha}=z_{\alpha}+n^{-1/2}p_{11}(z_{\alpha})+n^{-1}p_{21}(z_{\alpha})+O(n^{-3/2})
\]
where $p_{11}(x)=-p_{1}(x)$ and $p_{21}(x)=p_{1}(x)p'_{1}(x)-\frac{1}{2}x p_{1}^2(x)-p_{2}(x)$ with $p_{1}$ and $p_{2}$ defined by Equations~\ref{eqp_{1}} and~\ref{eqp_{2}}. 

Even though the true values of the parameter $\theta_0$, whose value may be required to compute the polynomials $p_1$ and $p_2$ in the Cornish-Fisher expansion when working with non pivotal statistics, are unknown, \cite{hall92} and \cite{jd96} showed that this issue can be solved using bootstrap techniques to ensure the interval still features the second-order probability matching property.

The same remark as above applies, using \cite{with83}, it is easy to derive Cornish-Fisher expansions of any higher order and complete our worksheet to have {\tt Maple} compute them.

\subsection{BCA bootstrap}\label{bcabootstrap}
In a framework valid for second-order Edgeworth expansions, bootstrap bias corrected and accelerated  (BCA) are second-order probability matching intervals \citep{hall92,jd96}. Yet, even though the simpler bootstrap $t$ intervals also feature second-order probability matching, BCA ones are often preferred \citep[p.~93 and pp.~133-134]{hall92} and their definition will be now shortly recalled. Note that the BCA intervals are transformation respecting and thus invariant under re-parametrization.

Let $\chi$ be a random sample from the population studied and $\theta$ a parameter of interest. Define $\hat H(x)=\Pit{\hat\theta^*\leqslant x|\chi}$ for the bootstrap distribution of $\hat\theta^*$ and, for all $\alpha$ in $]0;1[$, $\hat H^{-1}(\alpha)=\inf\{x:\hat H(x)\geqslant \alpha\}$. In this notation, $\hat y_{\alpha}=\hat H^{-1}(\alpha)=\hat H^{-1}(\Phi(z_\alpha))$. The sequences of quantiles $\{\hat y_{\alpha},\ 0<\alpha<1\}$ would be ``centered'' empirically if $\hat y_{1/2}$ were equal to $\hat\theta$. In general, this identity does not hold, and we might wish to correct for centering error. Let $\hat m=\Phi^{-1}\{\hat H(\hat\theta)\}$ be the estimate of the centering correction or bias correction. An one-sided BCA interval for $\theta_0$, with nominal coverage $1-\alpha$, is given by $(-\infty,\hat y_{BCA,1-\alpha})$ where:
\[
\hat y_{BCA,1-\alpha}=\hat H^{-1}(\Phi[\hat m+(\hat m+z_{1-\alpha})(1-\hat a(\hat m+z_{1-\alpha}))^{-1}]).
\]
The quantity $(1-\hat a(\hat m+z_{1-\alpha}))^{-1}$ amounts for a skewness correction. An explicit formula for the acceleration constant $a$, valid for non-parametric as well as parametric bootstrap, is worked out by Hall \cite[p.~132]{hall92}:
\[
a=\frac{A}{6\sigma^{3}\sqrt{n}},
\textup{ where } 
A=\sum_{i=1}^d\sum_{j=1}^d\sum_{k=1}^d a_ia_ja_k\mu_{ijk}.
\]
An estimate is also proposed by Hall \cite[p.~133]{hall92}:
\begin{eqnarray}
\hat a=\frac{\hat A}{6\hat \sigma^{3}\sqrt{n}},
\textup{ where } 
\hat A=\sum_{i=1}^d\sum_{j=1}^d\sum_{k=1}^d \hat a_i \hat a_j \hat a_k\hat {\mu}_{ijk}
\nonumber
\end{eqnarray}
\begin{eqnarray}
\textup{ with } \hat a_{i}=\left(\partial/\partial x^{(i)}\right)A(\mathbf{X})|_{\bf{X}=\boldsymbol{\mu}}\mbox{ and } 
\hat \mu_{ijk}=\frac1n\sum_{l=1}^n\left[\left(\mathbf{X_l}-\boldsymbol{\mu}\right)^{(i)}\left(\mathbf{X_l}-\boldsymbol{\mu}\right)^{(j)}\left(\mathbf{X_l}-\boldsymbol{\mu}\right)^{(k)}\right].\label{defamubca}\nonumber
\end{eqnarray}

A two-sided BCA interval is given by $(\hat y_{BCA,\alpha/2},\hat y_{BCA,1-\alpha/2})$.

\section{Expanding the mean}\label{expandmean}
Let $\Gamma_1$ be the skewness and $\kappa_1$ the excess kurtosis. All the explicit results in this section were derived from the theoretical ones stated in Section~\ref{sect1} using {\tt Maple}. For this first application we will follow closely the the steps detailed in the file \url{SI_BertrandMaumy_Edgeworth_mean.mw}. 
\subsection{Setting the framework}
We begin with a restart statement to be sure to start with a fresh session.
\begin{verbatim}
> restart;
\end{verbatim}
We first need to set within {\tt Maple} the framework of the Exemple~\ref{exempl1} in Section~\ref{subsec12}. We want to highlight that the user only needs to define the $g$ function, see Section~\ref{subsec12}, all the other quantities required for the computation, for both the non-studentized or studentized cases, been automatically derived from this very definition as soon as the previous function remained called $g$. 
\begin{verbatim}
> g:=proc(x1) return x1; end proc;
\end{verbatim}
\[\displaystyle g\, := \,\textbf{proc} (x1) \;\; \textbf{return}\,x1 \;\; \textbf{end\ proc};\]
Automatically retrieve the value of the dimension $d$, see Section~\ref{subsec12}, for the non-studentized case ({\tt Dim:=1}) and for the studentized one ({\tt Dims:=2}). Enlarge the vector to take into account higher moments required for the computations related to the studentized statistic.
\begin{verbatim}
> Dim:=nops([op(1,op(1,g))]);Dims:=2*Dim;NtoL:=[evaln(x||(1 .. Dims))];
\end{verbatim}
\[\displaystyle {\it Dim}\, := \,1\]
\[\displaystyle {\it Dims}\, := \,2\]
\[\displaystyle {\it NtoL}\, := \,[{\it x1},{\it x2}]\]
Even the $h$ function is automatically defined by this code chunk.
\begin{verbatim}
> h := proc (x::(seq(name))) option remember; local NNN, derg; 
	for NNN in seq(1..(_npassed/2)) do 
		derg[NNN]:=diff(g(op([1..Dim],[_passed])), 
		[op(NNN,[_passed])]);
	end do;
  return sqrt(sum(sum(derg[iii]*derg[jjj]*(op(iii+jjj,[_passed])-op(iii,
  [_passed])*op(jjj,[_passed])),jjj=1..(_npassed/2)),iii=1..(_npassed/2))); 
  end proc;
\end{verbatim}
\begin{eqnarray*}
&&\displaystyle h\, := \,\textbf{proc} (x::\mathit{seq} (name)) \\&&
\textbf{local} \,NNN,\,derg; \\&&
\textbf{option} \,remember; \\&&
\quad \textbf{for} \,NNN \,\textbf{in} \,\mathit{seq} (1..1/2 \ast \textbf{nargs}) \,\textbf{do}\\&&
\qquad derg[NNN]\,:=\,\mathit{diff} (\mathit{g} (\mathit{op} ([1..Dim],\,[\textbf{args}])),\,[\mathit{op} (NNN,\,[\textbf{args}])])\\&&
\quad \textbf{end\ do};\\&&
\quad \textbf{return}\,\mathit{sqrt} (\mathit{sum} (\mathit{sum} (derg[iii] \ast derg[jjj] \ast (\mathit{op} (iii + jjj,\,[\textbf{args}])-\mathit{op} (iii,\,[\textbf{args}])\\&&
\qquad \ast \mathit{op} (jjj,\,[\textbf{args}])),\,jjj \, = \, 1..1/2 \ast \textbf{nargs}),\,iii \, = \, 1..1/2 \ast \textbf{nargs}))\\&&
\textbf{end\ proc};
\end{eqnarray*}
We begin with the link between raw moments, from first to fourth order, and centered ones. We assume $\mu_3=\Gamma_1$ is the skewness, $\mu_4=\kappa_1+3$ the excess kurtosis and $\mu_i=E((X-\mu)^i)/\sigma^i$, $i \geqslant 5$, standardized central moments. Can be tuned with a specific distribution. The variable names $\Gamma_1$ and $\kappa_1$ had to be used to avoid errors due to confusion with two built-in {\tt R} functions {\tt Gamma} and {\tt kappa} if the result is exported to the {\tt R} statistical language.
\begin{verbatim}
> EspX:=[]:for ii in [seq(1..nops(NtoL))] 
  do EspX:=[op(EspX),cat(x,ii)=expand(sum(binomial(ii, j)*'mu||j'*
  sigma^j*mu^(ii-j), 'j' = 0 .. ii))]: end do:
  EspX;
\end{verbatim}
\[\displaystyle [{\it x1}=\mu,{\it x2}={\mu}^{2}+{\sigma}^{2}]\]
We reflect some known facts about the raw moments so that Maple can use them when simplifying complex expressions.
\begin{verbatim}
> for AA in [seq(1..nops(NtoL))] do if `mod`(AA,2)=1 then 
  assume(NtoL[AA],real) else assume(NtoL[AA]>0) end if end do;
  NtoL;
\end{verbatim}
\[\displaystyle [{\it x1},{\it x2}]\]
Several code chunks, not shown, now define various quantities of interests either for the computation itself of for speeding it up, namely: $SigmaA=h(\boldsymbol{\mu})$, $A0$, its partial derivatives $partA0$, $As$, its partial derivatives $partAs$, see Section~\ref{subsec12}.
The last preparation step, is to compute all the moments that appear in Section~\ref {subsec12}'s formulas. We only report the first and the last of the five code chunks required to carry out this work.
\begin{verbatim}
> for ii from 1 to 4*Dims do XX||ii:=t^ii-expand(subs([mu||4 = kappa1+3, 
  mu||3 = Gamma1, mu||2 = 1, mu||1 = 0, mu||0 = 1], sum(binomial(ii, j)*
  'mu||j'*sigma^j*mu^(ii-j), 'j' = 0 .. ii))): end do:
\end{verbatim}
\begin{verbatim}
> for ii from 1 to Dims do for jj from 1 to Dims do for kk from 1 to Dims 
  do for ll from 1 to Dims do Mu||ii||"."||jj||"."||kk||"."||ll:=simplify(
  subs([seq(t^(4*Dims-i)=t^(4*Dims-i)-XX||(4*Dims-i),i=0..(4*Dims-1))],
  expand(XX||ii*XX||jj*XX||kk*XX||ll))); end do; end do; end do; end do;
\end{verbatim}
\begin{verbatim}
> ExpGaussian:=[Gamma1=0,kappa1=0]: for ii from 5 to 4*Dims do ExpGaussian:=
  [op(ExpGaussian),mu||ii=(`mod`(ii+1, 2))*doublefactorial(ii-1)]: end do: 
  ExpGaussian;
\end{verbatim}
\[\displaystyle [{\it GAMMA1}=0,{\it kappa1}=0,{\it mu5}=0,{\it mu6}\\
\mbox{}=15,{\it mu7}=0,{\it mu8}=105]\]

\subsection{Acceleration constant $a$}
The purpose of the following code chunk is to compute  $A$ involved in the derivation of the value of the acceleration constant $a$, see Section~\ref{bcabootstrap}.
\begin{verbatim}
> A:=simplify(add(add(add(PartAs([NtoL[i]])*PartAs([NtoL[j]])*
  PartAs([NtoL[k]])*Mu||i||"."||j||"."||k,i=1..Dims),j=1..Dims),k=1..Dims)):
  A:=simplify(subs(EspX,A),assume=positive);
\end{verbatim}
\[\displaystyle A\, := \,\Gamma_1\]
As a consequence, we can state that the value of the acceleration constant $a$ is :
\[
a = \frac{\Gamma_1}{6\sqrt{n}}\cdot
\]
\subsection{Non-studentized statistic}
We now compute the values $k_{1,2}$, $k_{3,1}$, $k_{2,2}$ and $k_{4,1}$ defined in Theorem~\ref{T:5.1}. 
\begin{verbatim}
> k12:=simplify(1/2*(add(add(PartA0([NtoL[i],NtoL[j]])*Mu||i||"."||j,i=
  1..Dim),j=1..Dim))):
> k12:=simplify(subs(EspX,k12));
\end{verbatim}
\[\displaystyle {\it k12}\, := \,0\]
\begin{verbatim}
> k31:=k31:=simplify(add(add(add(PartA0([NtoL[i]])*PartA0([NtoL[j]])*PartA0(
  [NtoL[k]])*Mu||i||"."||j||"."||k,i=1..Dim),j=1..Dim),k=1..Dim)+3*add(add(
  add(add(PartA0([NtoL[i]])*PartA0([NtoL[j]])*PartA0([NtoL[k],NtoL[l]])*
  Mu||i||"."||k*Mu||j||"."||l,i=1..Dim),j=1..Dim),k=1..Dim),l=1..Dim)):
> k31:=simplify(subs(EspX,k31));
\end{verbatim}
\[\displaystyle {\it k31}\, := \,\Gamma_1\]
It is easy to derive results for a given distribution by replacing moments with their actual values. For instance, in the Gaussian case $\Gamma1=0$ and $\kappa1=0$. In the following, we will report only the results for the general case. 
\begin{verbatim}
> eval(k31,[Gamma1=0,kappa1=0]);
\end{verbatim}
\[\displaystyle 0\]
\begin{verbatim}
> k22:=simplify(add(add(add(PartA0([NtoL[i]])*PartA0([NtoL[j],NtoL[k]])*
  Mu||i||"."||j||"."||k,i=1..Dim),j=1..Dim),k=1..Dim)+1/2*add(add(add(add(
  PartA0([NtoL[i],NtoL[j]])*PartA0([NtoL[k],NtoL[l]])*Mu||i||"."||k*
  Mu||j||"."||l,i=1..Dim),j=1..Dim),k=1..Dim),l=1..Dim)+add(add(add(add(
  PartA0([NtoL[i]])*PartA0([NtoL[j],NtoL[k],NtoL[l]])*Mu||i||"."||j*
  Mu||k||"."||l,i=1..Dim),j=1..Dim),k=1..Dim),l=1..Dim)):
> k22:=subs(EspX,k22);
\end{verbatim}
\[\displaystyle {\it k22}\, := \,0\]
\begin{verbatim}
> k41:=simplify(add(add(add(add(PartA0([NtoL[i]])*PartA0([NtoL[j]])*PartA0(
  [NtoL[k]])*PartA0([NtoL[l]])*(Mu||i||"."||j||"."||k||"."||l-3*
  Mu||i||"."||j*Mu||k||"."||l),i=1..Dim),j=1..Dim),k=1..Dim),l=1..Dim)+12*
  add(add(add(add(add(PartA0([NtoL[i]])*PartA0([NtoL[j]])*PartA0([NtoL[k]])*
  PartA0([NtoL[l],NtoL[m]])*Mu||i||"."||l*Mu||j||"."||k||"."||m,i=1..Dim),
  j=1..Dim),k=1..Dim),l=1..Dim),m=1..Dim)+12*add(add(add(add(add(add(PartA0(
  [NtoL[i]])*PartA0([NtoL[j]])*PartA0([NtoL[k],NtoL[l]])*PartA0([NtoL[m],
  NtoL[o]])*(Mu||i||"."||k*Mu||j||"."||m*Mu||l||"."||o),i=1..Dim),j=1..Dim),
  k=1..Dim),l=1..Dim),m=1..Dim),o=1..Dim)+2/3*add(add(add(add(add(add(PartA0(
  [NtoL[i]])*PartA0([NtoL[j]])*PartA0([NtoL[k]])*PartA0([NtoL[l],NtoL[m],
  NtoL[o]])*(6*Mu||i||"."||m*Mu||k||"."||l*Mu||j||"."||o),i=1..Dim),
  j=1..Dim),k=1..Dim),l=1..Dim),m=1..Dim),o=1..Dim)):
> k41:=subs(EspX,k41);
\end{verbatim}
\[\displaystyle {\it k41}\, := \,\kappa_1\]
%
In order to compute $p_1$ and $p_2$, we now evaluate Equations~\ref{eqp_{1}} and~\ref{eqp_{2}}.
\begin{verbatim}
> P1:=-(k12+1/6*k31*(x^2-1)):
> P1:=simplify(P1);
\end{verbatim}
\[\displaystyle {\it P1}\, := \,-1/6\, \Gamma1 \, \left( {x}^{2}-1 \right) \]
\begin{verbatim}
> P2:=-x*(1/2*(k22+k12^2)+1/24*(k41+4*k12*k31)*(x^2-3)+1/72*k31^2*
  (x^4-10*x^2+15)):
> P2:=collect(collect(collect(P2,x),Gamma1),kappa1);
\end{verbatim}
\[\displaystyle {\it P2}\, := \, \left( -1/24\,{x}^{3}+1/8\,x \right) \kappa1+ \left( -{1}/{72}\,{x}^{5}+{5}/{36}\,{x}^{3}-{5}/{24}\,x\\
\mbox{} \right) {\Gamma1 }^{2}\]
\begin{verbatim}
> P11:=-P1:
> P11:=collect(collect(collect(simplify(subs(EspX,P11)),x),Gamma1),kappa1);
\end{verbatim}
\[\displaystyle {\it P11}\, := \, \left( 1/6\,{x}^{2}-1/6 \right)  \Gamma_1 \]
\begin{verbatim}
> P21:=P1*diff(P1,x)-P2-x*P1*P1/2:
> P21:=collect(collect(collect(simplify(subs(EspX,P21)),x),Gamma1),kappa1);
\end{verbatim}
\[\displaystyle  \left( 1/24\,{x}^{3}-1/8\,x \right) \kappa_1+ \left( -1/18\,{x}^{3}+{5}/{36}\,x \right) {\Gamma_1}^{2}\]

\subsection{Studentized statistic}\label{stumeansubsec}
\[
k_{1,2,s} = -\frac12\Gamma_1, \quad   
k_{3,1,s} = -2\Gamma_1, \quad   
k_{2,2,s} = 3+\frac74\Gamma_1^2, \quad
k_{4,1,s} = 6-2\kappa_1+12\Gamma_1^2.\nonumber
\]

\[
p_{1s}(x) = \frac{1}6\left(2x^2+1\right)\Gamma_1, \quad
p_{2s}(x) = \left(\frac1{12}x^3-\frac14x\right)\kappa_1+\left(-\frac1{18}x^5-\frac19x^3+\frac16x\right)\Gamma_1^2-\frac14x^3-\frac34x.\nonumber
\]
Same value as Shao and Tu \cite[p.~145]{jd96}.

We now provide the expansion of the $t$-distribution, \textit{cf}. remark~\ref{rematdist}. See the worksheet on how we used the polynomials $p_{1s}$ and $p_{2s}$ and an asymptotic expansion of the standard Gaussian CDF, density and polynomials $p_{1s}$ and $p_{2s}$ to get the polynomials $p_{1s,t\textup{-}dist}$ and $p_{2s,t\textup{-}dist}$ of the second-order Edgeworth expansion of the $t$-distribution.
\[
p_{1s,t\textup{-}dist} = p_{1s}(x), \quad
p_{2s,t\textup{-}dist} = p_{2s}(x)+\frac{1}2x.\nonumber
\]
These values are the one given by Lehman and Romano in Formulas 11.73 and 11.74 of Theorem 11.4.2, \cite[p.~460]{lero} for the second-order Edgeworth expansion of the $t$-distribution.
\[
p_{2s,t\textup{-}dist}(x) = \left(\frac1{12}x^3-\frac14x\right)\kappa_1+\left(-\frac1{18}x^5-\frac19x^3+\frac16x\right)\Gamma_1^2-\frac14x^3-\frac14x.\nonumber
\]
\[
p_{11s}(x) = -\frac16\left(2x^2+1\right)\Gamma_1, \quad
p_{21s}(x) = \left(-\frac1{12}x^3+\frac14x\right)\kappa_1+\left(\frac5{18}x^3-\frac5{72}x\right)\Gamma_1^2+\frac14x^3+\frac34x.
\]
Corrects the value of $p_{21s}$ given in Theorem 13.5 in \cite[p.~194]{dasg}. Same value as the one found by Shao and Tu in \cite[p.~145]{jd96} and Finner and Dickhaus in \cite{fd10}.

\subsection[{Exporting results from Maple to R}]{Export results from {\tt Maple} to {\tt R}}
\begin{verbatim}
> fd := fopen("A_mean.R", WRITE, TEXT);
\end{verbatim}
\[\displaystyle {\it fd}\, := \,2\]
\begin{verbatim}
> cg1 := CodeGeneration['Matlab'](A, resultname = 'A', output = string);
\end{verbatim}
\[\displaystyle {\it cg1}\, := \,``A = Gamma1;
\text{''}\]
\begin{verbatim}
> fprintf(fd, cg1);
\end{verbatim}
\[\displaystyle 12\]
\begin{verbatim}
> fclose(fd);
\end{verbatim}

\section{Expanding the variance}\label{expvarian}
All the explicit results in this section were derived from the theoretical ones stated in Section~\ref{sect1} using {\tt Maple}. The steps and details of the computations are available in the file \url{SI_BertrandMaumy_Edgeworth_variance.mw}. To demonstrate how is to simple to switch from the mean expansion in Section~\ref{expandmean} setting to another one, namely the variance expansion one, the only code chunk from that must be modified is the following one
\begin{verbatim}
> g:=proc(x1,x2) return x2-x1^2; end proc;
\end{verbatim}
\[\displaystyle g\, := \,\textbf{proc} (x1,x2) \;\; \textbf{return}\,x2-x1\wedge 2 \;\; \textbf{end\ proc};\]
\subsection{Acceleration constant $a$}
\[
A = -\frac{-\mu_6+3\kappa_1+7}{(\kappa_1+2)^{3/2}}\cdot
\]
In the Gaussian case, we find the same value as Shao and Tu in \cite[p.~138]{jd96} by substituting $(\kappa1,\mu_6)$ with the well known values $(\kappa1=0,\mu_6=15)$.
\[
a = \frac1{6\sqrt{n}}\frac{\left(15-3\cdot 0-7\right)}{(0+2)^{3/2}} = \frac{\sqrt{2}}{3\sqrt{n}}\cdot
\]

\subsection{Non-studentized statistic}
\[
k_{1,2} = -1/\sqrt{\kappa_1+2}, \quad k_{3,1} = -\left(-\mu_{6}+3\kappa_1+7+6\Gamma_1^2\right)/(\kappa_1+2)^{3/2}.
\]
Same as the value provided by Hall in \cite[p.~75 and 76]{hall92}.
\[
k_{2,2} = -2(\kappa_1+1)/(\kappa_1+2), \quad k_{4,1} = \left(3-24\Gamma_1\mu_{5}-4\mu_{6}+\mu_{8}-3\kappa_1^2+96\Gamma_1^2-6\kappa_1\right)/(\kappa_1+2)^2.
\]
The values of $p_1$, $p_2$, $p_{11}$ and $p_{21}$ can be found in the dedicated {\tt Maple} worksheet.

\subsection{Studentized statistic}
\[
k_{1,2,s} = \frac12(\kappa_1+3-\mu_{6}+4\Gamma_1^2)/(\kappa_1+2)^{3/2}, \quad k_{3,1,s} = 2(-\mu_{6}+3\kappa_1+3\Gamma_1^2+7)/(\kappa_1+2)^{3/2}
\]
Same value as the one found by Hall in \cite[p.~76]{hall92}.
\begin{eqnarray}
k_{2,2,s} &=& \frac14(20\kappa_1^3+163\kappa^2+56\Gamma_1^2\kappa_1+32\Gamma_1\kappa_1\mu_{5}-38\mu_{6}\kappa_1+450\kappa_1-90\mu_{6}+7\mu_{6}^2\nonumber\\
&&+415+112\Gamma_1^4+168\Gamma_1^2+64\Gamma_1\mu_{5}-56\Gamma_1^2\mu_{6})/(\kappa_1+2)^3\nonumber.
\end{eqnarray}
\begin{eqnarray}
k_{4,1,s} &=& 2(6\kappa_1^3+84\kappa_1^2+297\kappa_1+24\Gamma_1\kappa_1\mu_{5}-32\mu_{6}\kappa_1+54\Gamma_1^2\kappa_1-\kappa_1\mu_{8}-\nonumber\\
&&2\mu_{8}+312+72\Gamma_1^4-42\Gamma_1^2\mu_{6}+6\mu_{6}^2+48\Gamma_1\mu_{5}+150\Gamma_1^2-76\mu_{6})/(\kappa_1+2)^3\nonumber.
\end{eqnarray}
The values of $p_{1,s}$, $p_{2,s}$, $p_{11,s}$ and $p_{21,s}$ can be found in the dedicated {\tt Maple} worksheet.

\section{A more complex example}\label{morecomp}
All the explicit results in this section were derived from the theoretical ones stated in Section~\ref{sect1} using {\tt Maple}. The steps and details of the computations are available in the file \url{SI_BertrandMaumy_Edgeworth_ML_normal_std_sym.mw}.

\subsection{Introduction}
One burning issue that commonly arises in official quality laboratories or in the industries of various sectors, namely chemistry, pharmacy, food processing, etc, is the estimation of the proportion of experimental results that lie within two limits. For instance, in official quality laboratories, $[L,U]$ is the the range of the acceptable values and $\pi$ is the proportion of valid units. 

Let $\{X_i: 1\leqslant i\leqslant n\}$, where $n \geqslant 2$, be a sequence of independent experimental results that follow a Gaussian distribution with mean $\mu$ and variance $\sigma^2>0$.
We denote the sample mean and the sample variance respectively by: 
\[
\begin{array}{lr}
\overline{X}=\displaystyle\frac{1}{n}\sum_{i=1}^nX_i\quad\quad& \displaystyle s^2_c=\displaystyle\frac{1}{n-1}\sum_{i=1}^n\left(X_i-\overline{X}\right)^2.
\end{array}
\]
 
We introduce the proportion $\pi$ defined by:
\begin{equation}\label{E:0.1}
\pi:=\pi\left(\mu,\sigma^2\right)=\Pit{L\leqslant X\leqslant U}
=\Phi\left(\frac{U-\mu}{\sigma}\right)-\Phi\left(\frac{L-\mu}{\sigma}\right),
\end{equation}
where $\Phi$ stands for the cumulative distribution function of the standard Gaussian distribution, 
$L$ and $U$, with $L<U$, are the ``limits of acceptance'', whose values depend on the active regulatory norm.

Using Equation~\ref{E:0.1}, one can put forward $\widehat{\pi}$ as the maximum likelihood (ML) point estimator of $\pi$:
\begin{equation}\label{E:2.3}
\widehat{\pi}:=\pi\left(\overline{X},\frac{s_c^2}{w^2}\right)=\Phi\left(w\,\frac{U-\overline{X}}{s_c}\right)
-\Phi\left(w\,\frac{L-\overline{X}}{s_c}\right),
\end{equation} 
where $w^2=n/(n-1)$.

\subsection{Towards the expansion}
Following Section~\ref{subsec12}, let's introduce the function non-studentized function $A_{{\scriptstyle{ML}},0}$ and the studentized function $A_{{\scriptstyle{ML}},s}$ defined by: 
\begin{eqnarray}
&&\begin{array}{cccccc}
\label{FML}&&A_{{\scriptstyle{ML}},0}:
(x,y)\in\mathbb{R}^2&\mapsto&\left(g_{{\scriptstyle{ML}}}(x,y)-g_{{\scriptstyle{ML}}}(\mu,\sigma^2+\mu^2)\right)/h({\boldsymbol \mu})\in\mathbb{R}\\
\end{array}
\\
&&\begin{array}{cccccc}
\label{FMLS}&&A_{{\scriptstyle{ML}},s}:
(x,y,z,t)\in\mathbb{R}^4&\mapsto&\left(g_{{\scriptstyle{ML}}}(x,y)-g_{{\scriptstyle{ML}}}(\mu,\sigma^2+\mu^2)\right)/h(x,y,z,t)\in\mathbb{R}\\
\end{array}\label{FsML}
\end{eqnarray}
where $g_{{\scriptstyle{ML}}}(x,y)=\pi\left(x,y-x^2\right)$, $\pi$ was defined in Equation~\ref{E:0.1}, ${\boldsymbol \mu}=(\mu,\mu^2+\sigma^2,\mu^3+3\mu\sigma^2,\mu^4+6\mu^2\sigma^2+3\sigma^4)$ and, with $\psi(k)$ the $k$th element of the vector ${\bf x}=(x,y,z,t)$, we have:
\[
h_{\scriptstyle{ML}}^2({\bf x})=\sum_{i=1}^2\sum_{j=1}^2\frac{\partial
{g_{{\scriptstyle{ML}}}}}{\partial {\bf x}_i}(x,y)\times \frac{\partial {g_{{\scriptstyle{ML}}}}}{
\partial {\bf x}_j}(x,y)\times\left(\psi(i+j)-\psi(i)\psi(j)\right).
\] 
Please note that  $g_{{\scriptstyle{ML}}}(\mu,\sigma^2+\mu^2)=\pi\left(\mu,\sigma^2\right)$.

\begin{verbatim}
> gML:=proc(x1,x2); return 1/sqrt(2*Pi)*int(exp(-t^2/2),t=(-lambda-x1)/
  sqrt(x2-x1^2)..(lambda-x1)/sqrt(x2-x1^2)); end proc;
\end{verbatim}
\begin{eqnarray*}
&& gML := \,\textbf{proc} (x1, x2) \\
&&\qquad  \textbf{return}\, \mathit{sqrt} (2 \ast Pi)\hat{~}{-1} \ast \mathit{int} (\mathit{exp} (-1/2 \ast t\hat{~}{2}),\,t \, = \, (-lambda-x1) \ast \\
&&\qquad \mathit{sqrt} (x2-x1\hat{~}{2})\hat{~}{-1}..(lambda-x1) \ast \mathit{sqrt} (x2-x1\hat{~}{2})\hat{~}{-1})\\
&&\textbf{end\ proc};
\end{eqnarray*}

\begin{corollary}\label{C:5.1}We have:
\[h_{\scriptstyle{ML}}({\boldsymbol \mu})=\frac{1}{2\pi}\left(\left(e^{-\frac{(U-\mu)^2}{2\sigma^2}}-e^{-\frac{(\mu-L)^2}{2\sigma^2}}\right)^2+\frac{1}{2}\left(\frac{U-\mu}{\sigma}e^{-\frac{(U-\mu)^2}{2\sigma^2}}+\frac{\mu-L}{\sigma}e^{-\frac{(\mu-L)^2}{2\sigma^2}}\right)^2\right).\]
\end{corollary}

Thanks to a direct computation \cite{mee88}, one can easily check that the value of $h_{\scriptstyle{ML}}({\boldsymbol \mu})$ is equal to the asymptotic variance of $n^{1/2}(\widehat{\pi}-\pi)$ given in corollary \ref{C:5.1}.

We need to adapt the definition of the moments to the Gaussian frame.
\begin{verbatim}
> mu[0]:=1: mu[1]:=0: mu[2]:= 1: mu[3]:=Gamma1: Gamma1:=0: mu[4]:=kappa1+3: 
  kappa1:=0: for ii from 5 to 4*Dims do mu[ii]:=`mod`(ii+1,2)*
  doublefactorial(ii-1); end do:
> EspX:=[]:for ii in [seq(1..nops(NtoL))] do EspX:=[op(EspX),cat(x,ii)=
  expand(sum(binomial(ii, j)*mu[j]*sigma^j*mu^(ii-j), j = 0 .. ii))]: 
  end do:EspX;
\end{verbatim}
\[\displaystyle [{\it x1}=\mu,{\it x2}={\mu}^{2}+{\sigma}^{2},{\it x3}={\mu}^{3}+3\,{\sigma}^{2} \\
\mbox{},{\it x4}={\mu}^{4}+6\,{\sigma}^{2}{\mu}^{2}+3\,{\sigma}^{4}]\]

%

\subsection[Expanding the estimator pi hat]{Expanding the estimator $\widehat{\pi}$}
\begin{theorem}\label{T:5.1bis}
We have the following second-order Edgeworth expansion:
\begin{eqnarray}
\label{E:4.7bis}
\Pit{\sqrt n F_{{\scriptstyle{ML}},\bullet}\left(\overline{X},s_c^2/w^2\right)\leqslant x}
&=&
\Phi(x)+n^{-1/2}p_{1,{\scriptstyle{ML}},\bullet}(x)\phi(x)\\
&&
+n^{-1}p_{2,{\scriptstyle{ML}},\bullet}(x)\phi(x)+
\nonumber
\,O\left(n^{-3/2}\right),
\end{eqnarray}
uniformly in $x\in\mathbb{R}$ and where $p_{1,{\scriptstyle{ML}},\bullet}$ and $p_{2,{\scriptstyle{ML}},\bullet}$ are defined by Equations~\ref{E:4.1bis} and~\ref{E:4.2bis} and $\phi$ is the probability density function of a standard Gaussian distribution. Moreover, we have:
\begin{eqnarray}
p_{1,{\scriptstyle ML},\bullet}(x)&=&-\left(k_{1,2,{\scriptstyle{ML}},\bullet}+\frac{1}{6}k_{3,1,{\scriptstyle{ML}},\bullet}(x^2-1)\right),\label{E:4.1bis}\\
p_{2,{\scriptstyle ML},\bullet}(x)&=&-x\left(\frac{1}{2}\left(k_{2,2,{\scriptstyle{ML}},\bullet}+k^2_{1,2,{\scriptstyle{ML}},\bullet}\right)
+\frac{1}{24}\left(k_{4,1,{\scriptstyle{ML}},\bullet}+4k_{1,2,{\scriptstyle{ML}},\bullet}k_{3,1,{\scriptstyle{ML}},\bullet}\right)(x^2-3)+\right.\nonumber\\
&&\left.\frac{1}{72}k^2_{3,1,{\scriptstyle{ML}},\bullet}(x^4-10x^2+15)\right)\label{E:4.2bis}.
\end{eqnarray}
\end{theorem}
We assume, for the sake of simplifying the quantities displayed that $\mu=0$. Note that $\sigma$ remains unknown and let $\check \lambda=\lambda/\sigma$. In addition, as we stressed it before, in many regulatory issues, the limits of the interval of acceptance, $L$ and $U$, are opposite numbers: $U=\lambda=-L$.
\begin{proposition}[Symmetric setting]
We begin with the values for the non-studentized statistic:
\begin{eqnarray*}
\sigma_{{\scriptstyle{ML}}}^2&=&\frac{1}{\pi}{\check\lambda}^2e^{-{\check\lambda}^2}, \quad
k_{1,2,{\scriptstyle{ML}},0}=\frac{1}{2\sqrt{2}}(3-{\check\lambda}^2), \quad k_{2,2,{\scriptstyle{ML}},0}=\frac{3}{4}(5-6{\check\lambda}^2+{\check\lambda}^4),\\ k_{3,1,{\scriptstyle{ML}},0}&=&\frac{1}{\sqrt{2}}(5-3{\check\lambda}^2),\quad
k_{4,1,{\scriptstyle{ML}},0}=24-32{\check\lambda}^2+8{\check\lambda}^4.
\end{eqnarray*}
Then we follow with the values for the studentized statistic:
\begin{eqnarray*}
k_{1,2,{\scriptstyle{ML}},s}&=&\frac{1}{2\sqrt{2}}(1+{\check\lambda}^2), \quad k_{2,2,{\scriptstyle{ML}},s}=\frac{1}{4}(35+10{\check\lambda}^2+3{\check\lambda}^4), \quad k_{3,1,{\scriptstyle{ML}},s}=\frac{1}{\sqrt{2}}(-1+3{\check\lambda}^2),\\
k_{4,1,{\scriptstyle{ML}},s}&=&18+4{\check\lambda}^2+8{\check\lambda}^4.
\end{eqnarray*}
Hence, the second-order Edgeworth expansion of $F_{{\scriptstyle{ML}},0}$ stated in Equation~\ref{E:4.7bis} boils down to, uniformly in $x\in\mathbb{R}$:
\begin{eqnarray}
\Pit{\sqrt n F_{{\scriptstyle{ML}},0}\left(\overline{X},s_c^2/w^2\right)\leqslant x}
&=&
\Phi(x)+\frac{1}{\sqrt{n}}\left(-4+(-5+3{\check\lambda}^2)x^2\right)\frac{\phi(x)}{6\sqrt{2}}+\frac{1}{n}\Big[-x+\left(\frac{2}{3}-\right.\nonumber\\
&&{\check\lambda}^2+{\check\lambda}^4\bigg)x^3+\left(-\frac{25}{24}+\frac{5}{4}{\check\lambda}^2-\frac{3}{8}{\check\lambda}^4\right)x^5\Big]\frac{\phi(x)}{6}+\,O\left(n^{-3/2}\right).\nonumber
\end{eqnarray}
As to $F_{{\scriptstyle{ML}},s}$, its second-order Edgeworth expansion stated in Equation~\ref{E:4.7bis} boils down to, uniformly in $x\in\mathbb{R}$:
\begin{eqnarray}
\Pit{\sqrt n F_{{\scriptstyle{ML}},s}\left(\overline{X},s_c^2/w^2\right)\leqslant x}
&=&
\Phi(x)+\frac{1}{\sqrt{n}}\left(-4+(1-3{\check\lambda}^2)x^2\right)\frac{\phi(x)}{6\sqrt{2}}+\frac{1}{n}\Big[-\frac{29}{2}x-\left(\frac{23}{6}+\right.\nonumber\\
&&4{\check\lambda}^2-{\check\lambda}^4\bigg)x^3+\left(-\frac{1}{24}+\frac{1}{4}{\check\lambda}^2-\frac3{8}{\check\lambda}^4\right)x^5\Big]\frac{\phi(x)}{6}+\,O\left(n^{-3/2}\right).\nonumber
\end{eqnarray}
These results lead to Cornish-Fisher expansions of the quantiles of the estimator $\widehat{\pi}$ given by the following equation:
\begin{eqnarray}\label{devcornfish}
w_{\alpha}^{0,{\scriptstyle{ML}}}&=&z_{\alpha}+\frac{1}{6\sqrt{2n}}\left(4+(5-3{\check\lambda}^2)z_{\alpha}^2\right)+\frac{z_{\alpha}}{36n}\left(22-12{\check\lambda}^2+(11-18{\check\lambda}^2+3{\check\lambda}^4)z_{\alpha}^2\right)+\nonumber\\
&&O\left(n^{-3/2}\right),\\
w_{\alpha}^{s,{\scriptstyle{ML}}}&=&z_{\alpha}+\frac{1}{6\sqrt{2n}}\left(4+(-1+3{\check\lambda}^2)z_{\alpha}^2\right)+\frac{z_{\alpha}}{36n}\left(79+12{\check\lambda}^2 +(26+12{\check\lambda}^2+3{\check\lambda}^4)z_{\alpha}^2\right)+\nonumber\\
&&O\left(n^{-3/2}\right),
\end{eqnarray}
where $z_{\alpha}$ is the $\alpha$-level standard Normal quantile.
\end{proposition}

\begin{proposition}[Symmetric setting]
With the same assumptions as above, the acceleration constant $a$ is:
\[
A=-2\sqrt2, \quad a=-\frac{\sqrt2}{3\sqrt{n}}\cdot
\]
\end{proposition}
Hence we showed that the value of the acceleration constant $a$ does not depend of the value of either $\sigma$ or $\lambda$, which could not be easily guessed from the expression of the $\hat \pi$ statistic. If $\mu\neq 0$, unfortunately this property does no longer hold. 

Nevertheless the value of the accelaration constant can be computed with {\tt Maple} and then exported into {\tt R} compatible code, which can be later sourced into an {\tt R} script, using the following code chunks.

\begin{verbatim}
> fd := fopen("AML_stdsym.R", WRITE, TEXT);
\end{verbatim}
\[\displaystyle {\it fd}\, := \,2\]
\begin{verbatim}
> cg1 := CodeGeneration['Matlab'](AML, resultname = 'Acent', output = string);
\end{verbatim}
\[\displaystyle {\it cg1}\, := \,``Acent = -0.2e1 * sqrt(0.2e1);
\text{''}
\]

\begin{verbatim}
> cg2 := CodeGeneration['Matlab'](AMLT, resultname = 'A', output = string);
\end{verbatim}
\begin{eqnarray*}
\displaystyle && cg2 := \,``A = -0.2e1 * sqrt(0.2e1) * (exp(-(mu + lambda) \wedge 2 / sigma \wedge 2 / 0.2e1) * mu +\\&&
 exp(-(mu +lambda) \wedge 2 / sigma \wedge 2 / 0.2e1) * lambda + exp(-(lambda - mu) \wedge 2 / \\&&
 sigma \wedge 2 / 0.2e1) * lambda - exp(-(lambda - mu) \wedge 2 / sigma \wedge 2 / 0.2e1) * mu) * (0.3e1 *\\&&
  exp(-(mu +lambda) \wedge 2 /  sigma \wedge 2) *sigma \wedge 2 + 0.3e1 * exp(-(lambda - mu) \wedge 2 \\&&
 / sigma\wedge 2) * sigma \wedge 2 + 0.2e1 * exp(-(mu + lambda) \wedge 2 /sigma \wedge 2) * mu * lambda \\&&
 - 0.2e1 *exp(-(mu \wedge 2 + lambda \wedge 2) / sigma \wedge 2) * mu \wedge 2 - 0.2e1 * exp(-(lambda \\&&
 - mu) \wedge 2 /sigma \wedge 2) * mu * lambda + 0.2e1 * exp(-(mu \wedge 2 + lambda \wedge 2) / sigma \wedge 2) \\&&
* lambda \wedge 2 +exp(-(mu + lambda) \wedge 2 / sigma \wedge 2) * mu \wedge 2 - 0.6e1 * exp(-(mu \wedge 2 \\&&
+ lambda \wedge 2) /sigma \wedge 2) * sigma \wedge 2 + exp(-(lambda - mu) \wedge 2 / sigma \wedge 2) * mu \wedge 2 \\&&
+ exp(-(mu +lambda) \wedge 2 / sigma \wedge 2) * lambda \wedge 2 + exp(-(lambda - mu) \wedge 2 /  \\&&
sigma \wedge 2) *lambda \wedge 2) * (exp(-(mu + lambda) \wedge 2 / sigma \wedge 2) * mu \wedge 2 - 0.2e1 \\&&
* exp(-(mu \wedge 2 +lambda \wedge 2) / sigma \wedge 2) * mu \wedge 2 + exp(-(lambda - mu) \wedge 2 / \\&&
sigma \wedge 2) * mu \wedge 2 - 0.4e1 * exp(-(mu \wedge 2 + lambda \wedge 2) / sigma \wedge 2) * sigma \wedge 2 \\&&
+ 0.2e1 * exp(-(mu + lambda) \wedge 2 / sigma \wedge 2) * sigma \wedge 2 + 0.2e1 * exp(-(lambda \\&& 
- mu) \wedge 2 /sigma \wedge 2) * sigma \wedge 2 + 0.2e1 * exp(-(mu + lambda) \wedge 2 / sigma \wedge 2) * mu \\&&
* lambda - 0.2e1 * exp(-(lambda - mu) \wedge 2 / sigma \wedge 2)  * mu* lambda + 0.2e1 * \\&&
exp(-(mu \wedge 2 + lambda \wedge 2) / sigma \wedge 2) * lambda \wedge 2 + exp(-(mu +lambda) \wedge 2 / \\&&
 sigma \wedge 2) * lambda \wedge 2  + exp(-(lambda - mu) \wedge 2 / sigma \wedge 2) * lambda \wedge 2) \wedge (\\&&
-0.3e1 / 0.2e1);\\&&
\text{''}
\end{eqnarray*}

\begin{verbatim}
> fprintf(fd, cg1);fprintf(fd, cg2);
\end{verbatim}
\[\displaystyle 34\]
\[\displaystyle 1495\]

\begin{verbatim}
> fclose(fd);
\end{verbatim}

\begin{figure}[tb]
\begin{tabular}{cc}
\includegraphics[angle=0,width=0.5\textwidth]{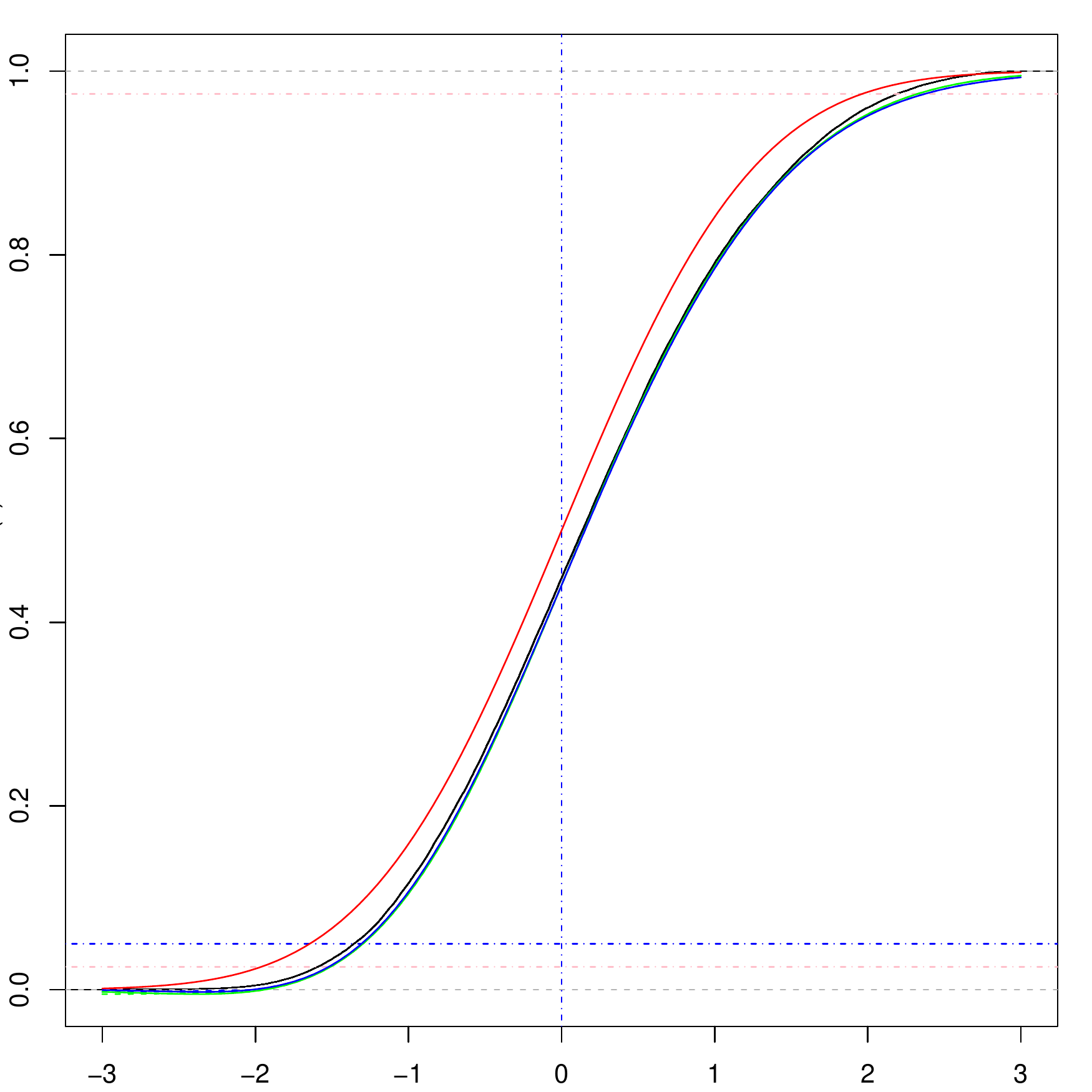}&
\includegraphics[angle=0,width=0.5\textwidth]{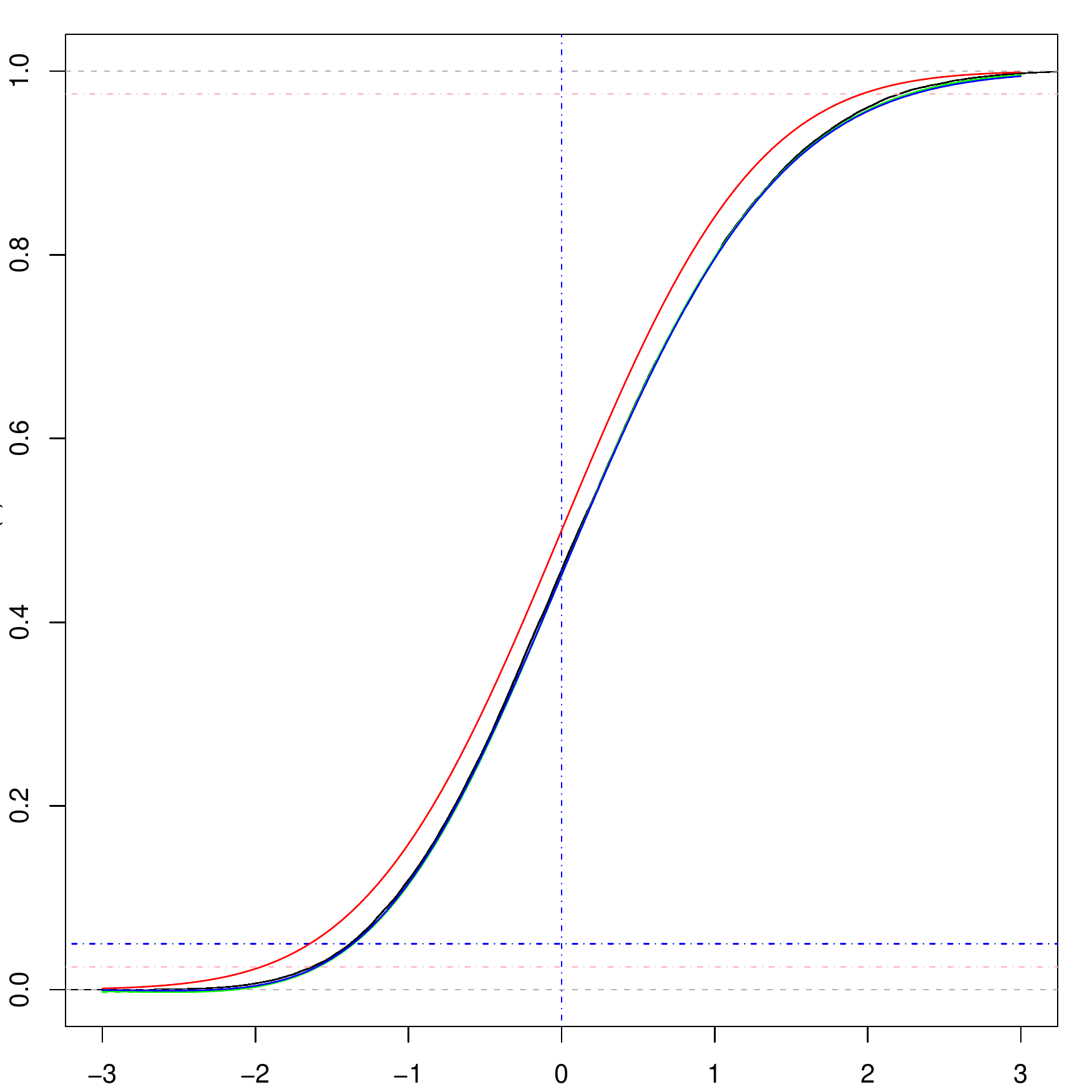}\\
\includegraphics[angle=0,width=0.5\textwidth]{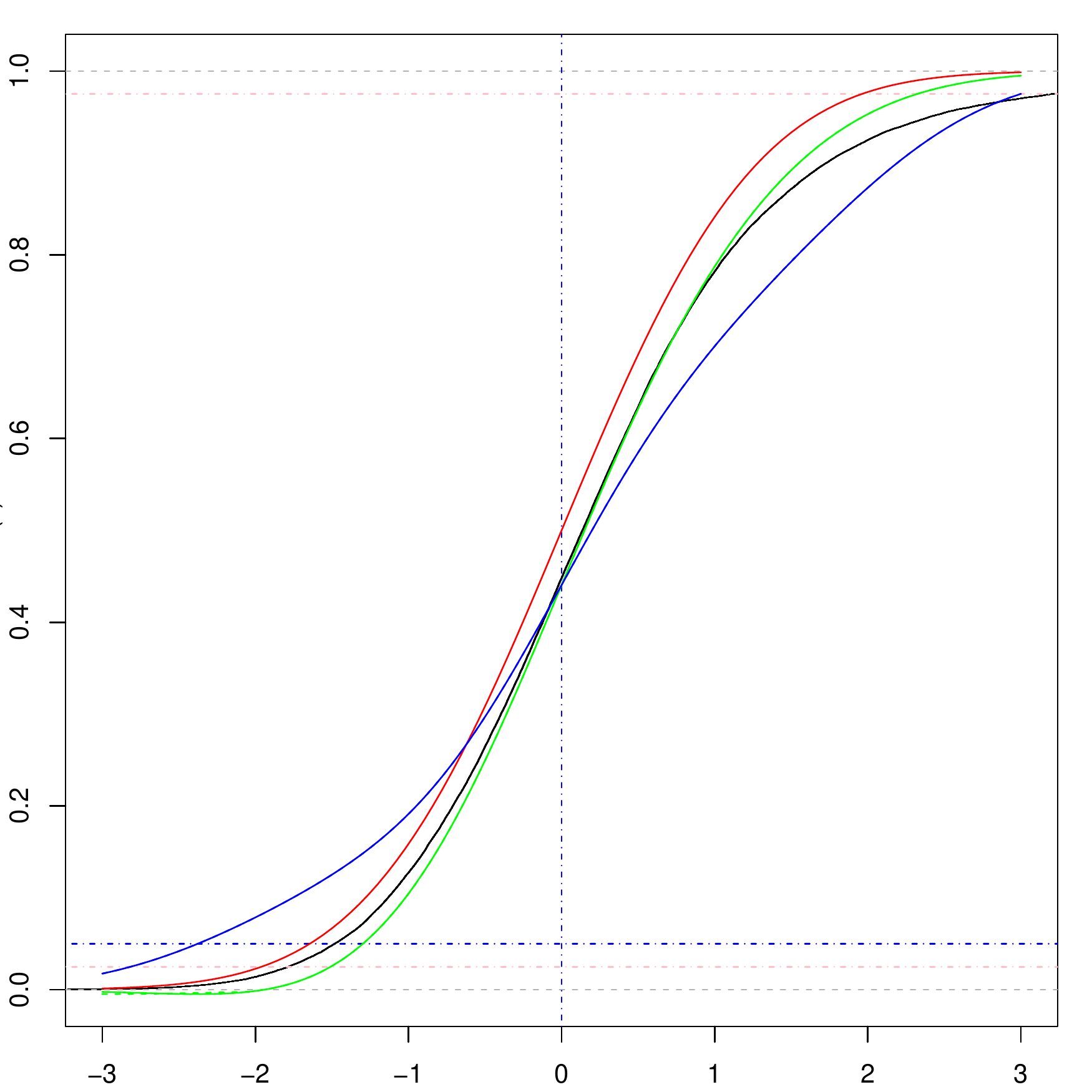}&
\includegraphics[angle=0,width=0.5\textwidth]{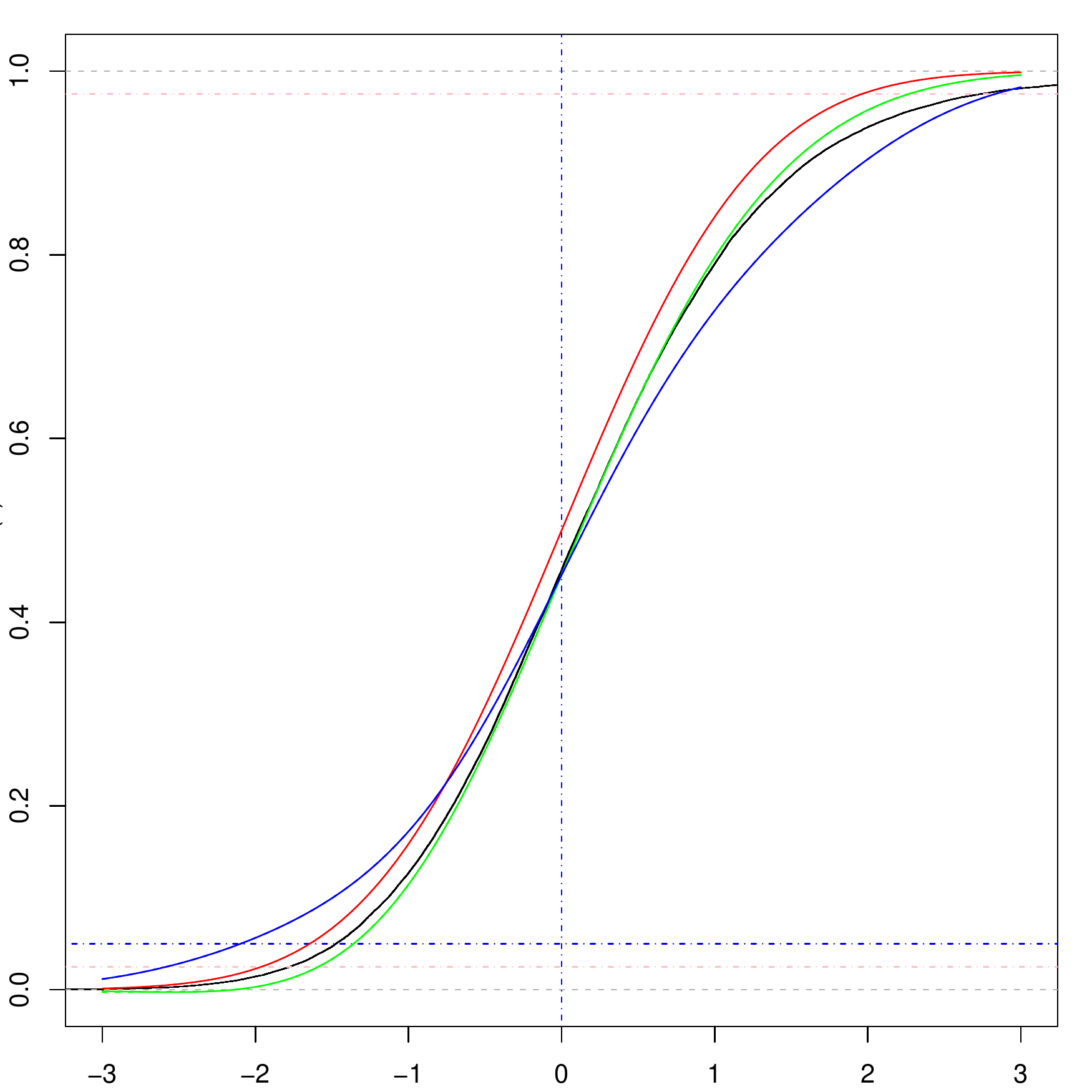}\\
\end{tabular}
\caption{Cumulative distribution functions for the non-studentized (up) and studentized (down) ML estimators for a sample size of 10 (left) and 15 (right), $\mu=0$, $\sigma=1$ and $\lambda=1$ thus $\pi\approx 0.68$.}
\label{1}
\end{figure}

\begin{figure}[tb]
\begin{tabular}{cc}
\includegraphics[angle=0,width=0.5\textwidth]{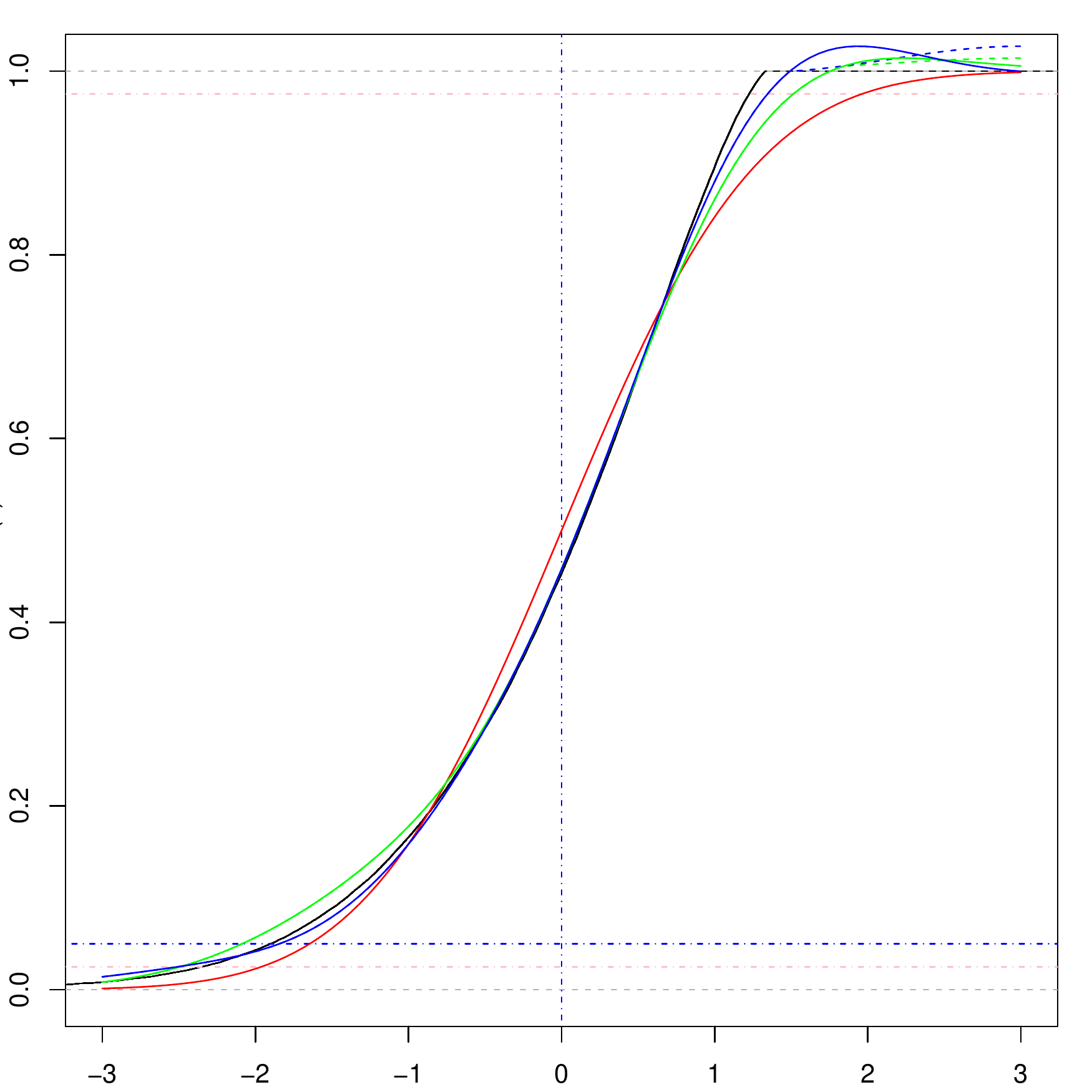}&
\includegraphics[angle=0,width=0.5\textwidth]{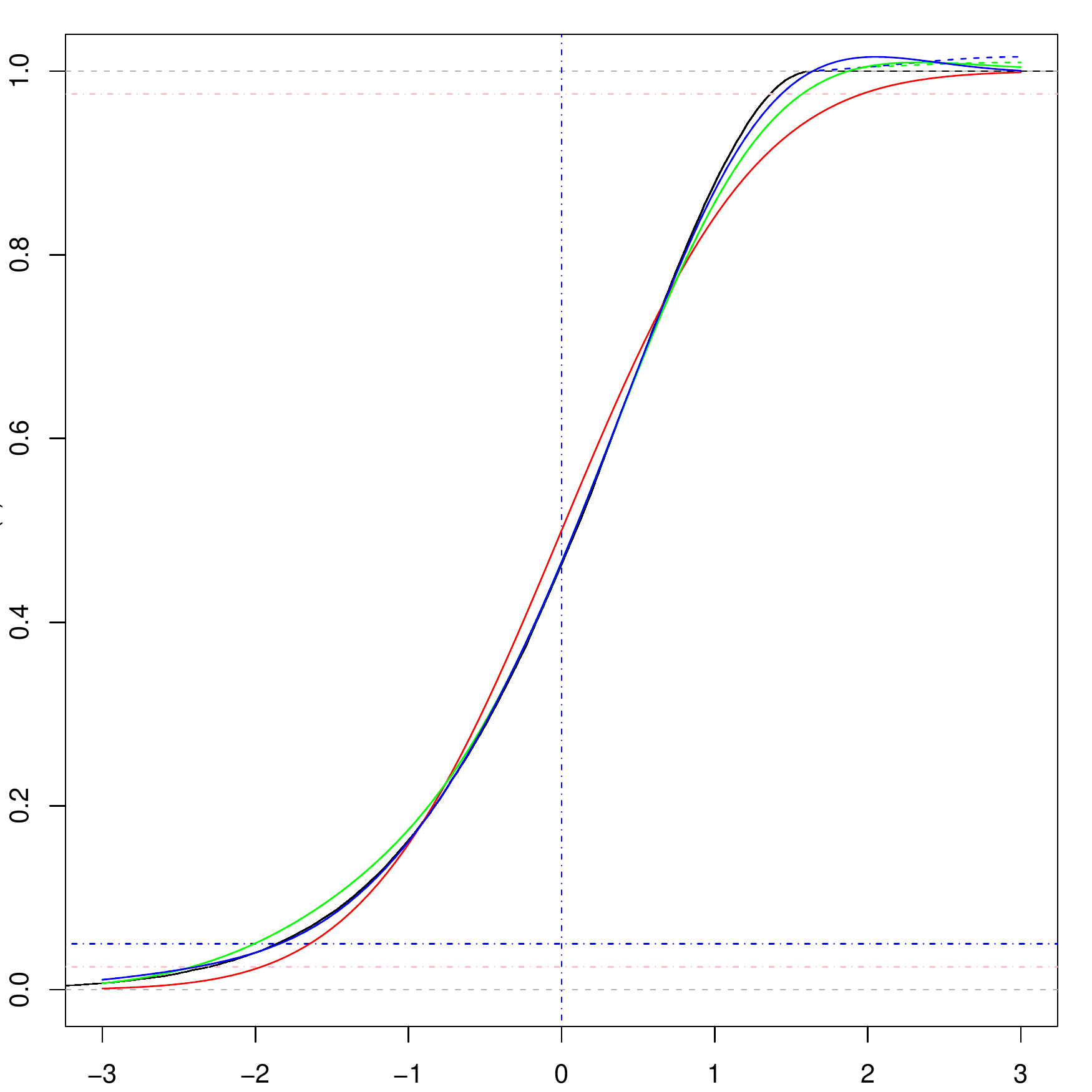}\\
\includegraphics[angle=0,width=0.5\textwidth]{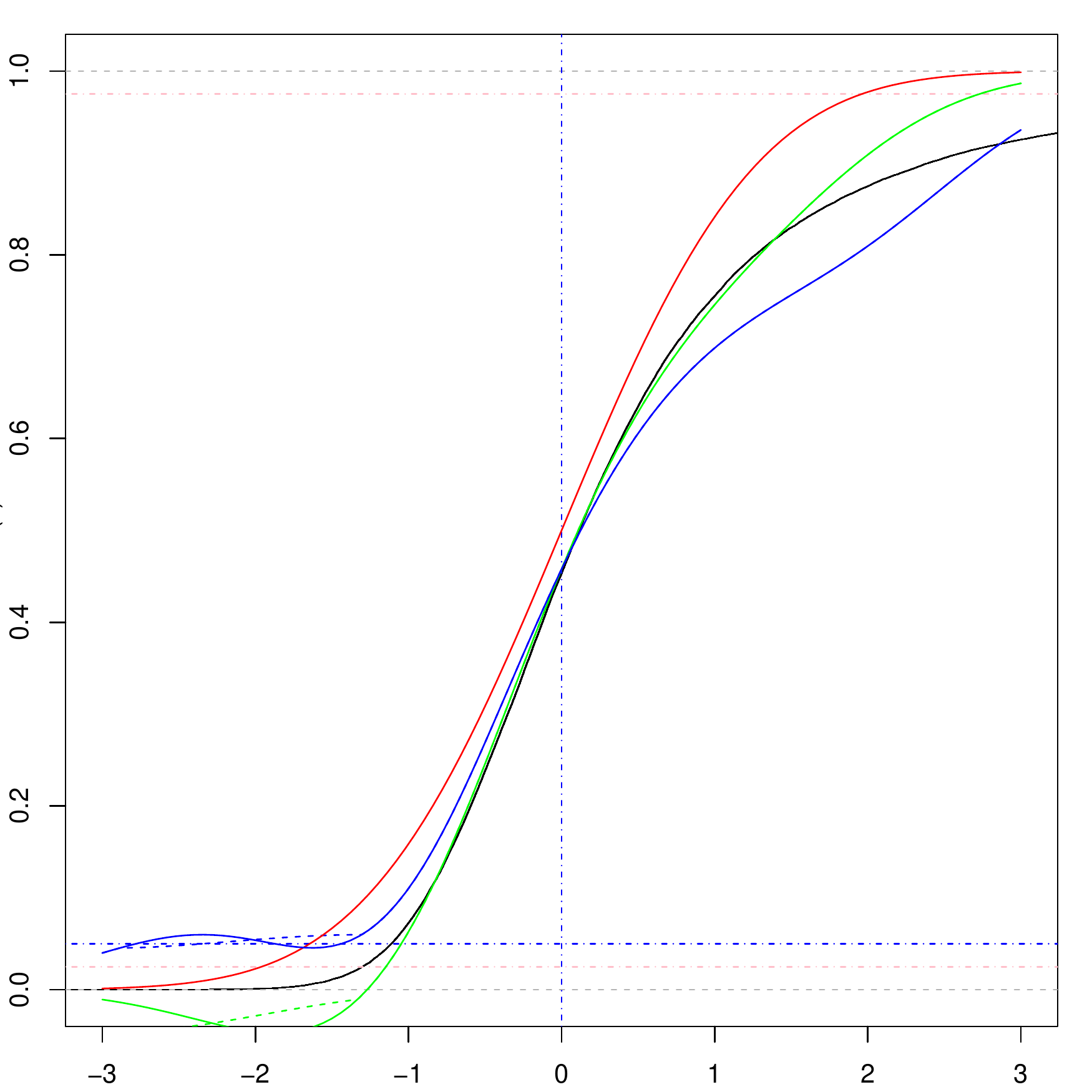}&
\includegraphics[angle=0,width=0.5\textwidth]{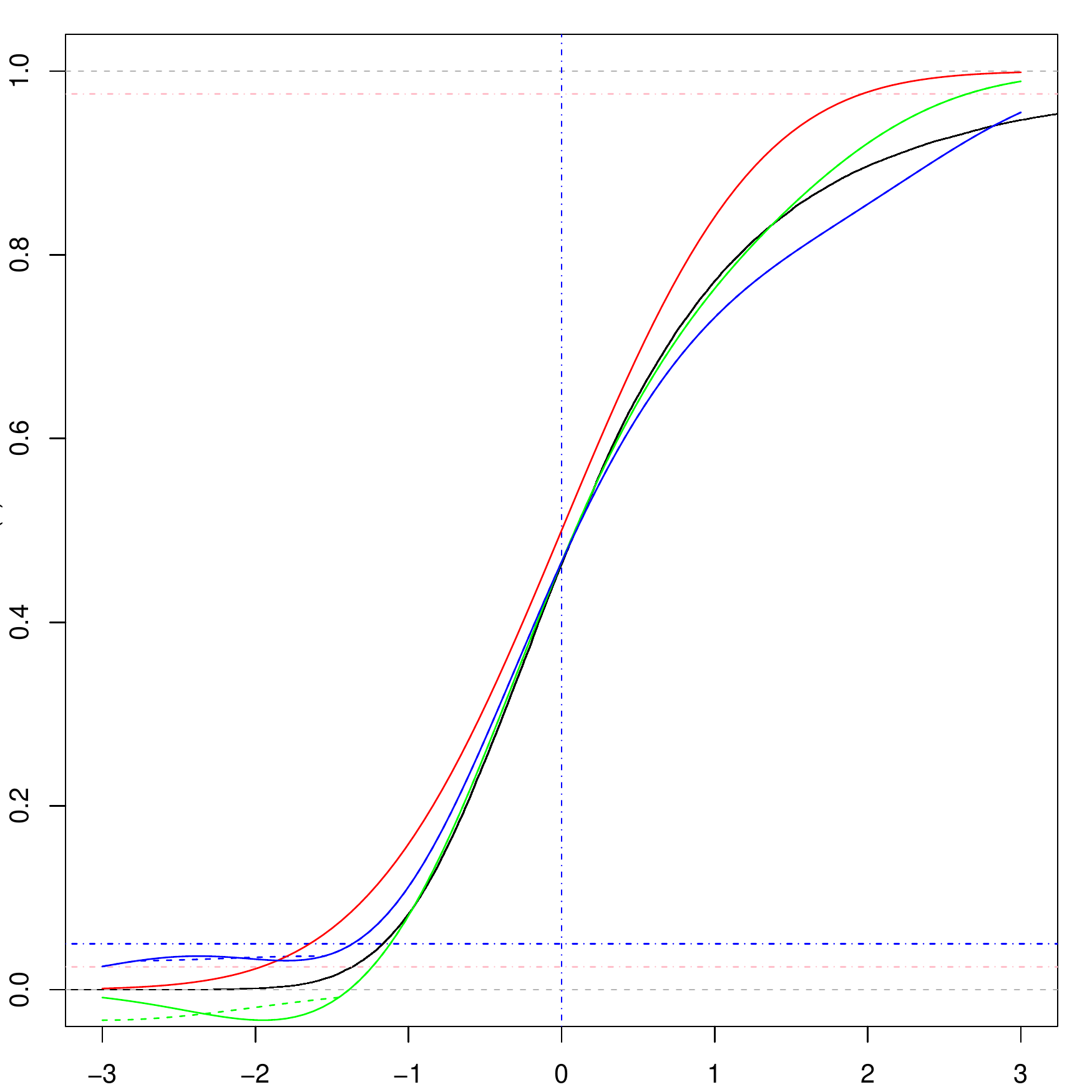}\\
\end{tabular}
\caption{Cumulative distribution functions for the non-studentized (up) and studentized (down) ML estimators for a sample size of 20 (left) and 30 (right), $\mu=0$, $\sigma=1$ and $\lambda=2$ thus $\pi\approx 0.95$.}
\label{2}
\end{figure}

We displayed on Figures~\ref{1} and~\ref{2} eight examples of approximations of the the real cumulative distribution functions (black) of the $\widehat{\pi}$ estimator. Non-increasing cumulative distribution functions resulting from the above Edgeworth expansions (see Figure~\ref{2} for blatant examples) were dealt with using rearrangement \citep{cfg10,rearrR} and plotted with dash style (- -) whereas the original cumulative distribution functions were plotted with plain style (---). In all the simulations where it is useful, rearrangement improves the approximation of the Edgeworth expansions to the true cumulative distribution functions.

In the case of the non-studentized statistics, the second-order Edgeworth expansion (blue) of the cumulative distribution function greatly improves the whole approximation to the real cumulative distribution functions over the normal approximation (red) and even over the first order Edgeworth one (green). In this non-studentized case, and even though Edgeworth expansions are asymptotically valid, the second-order one performs nicely even for very small sample sizes.

As to the studentized statistics, the second-order Edgeworth expansion (blue) of the cumulative distribution function greatly improves the approximation to the upper quantiles (the 97,5\% quantile is shown by the upper pink dash-dot line) over the normal approximation (red) and even over the first order Edgeworth one (green). In this studentized case, and even though Edgeworth expansions are asymptotically valid, the second-order one performs nicely even for small sample sizes.

Our worksheet handles the general asymmetric setting, i.e., $L\neq -U$ and even if, in addition, $\mu$ or $\sigma$ are not assigned to specific values -see the file \url{SI_BertrandMaumy_Edgeworth_ML_normal_gen.mw}-. 


\bibliographystyle{plainnat}

\end{document}